\documentclass[10pt,conference]{IEEEtran}

\usepackage{amsmath}
\usepackage{bbold}
\usepackage[utf8]{inputenc}
\usepackage{graphicx}
\usepackage[draft=true]{hyperref}
\usepackage{verbatim}

\newtheorem{definition}{Definition}[section]

\DeclareMathOperator{\AllTests}{\mbox{\footnotesize AllTests}}
\DeclareMathOperator{\DependentTests}{\mbox{\footnotesize DependentTests}}
\DeclareMathOperator{\SelectedTests}{\mbox{\footnotesize SelectedTests}}
\DeclareMathOperator{\FailedTests}{\mbox{\footnotesize FailedTests}}
\DeclareMathOperator{\FlakedTests}{\mbox{\footnotesize FlakedTests}}
\DeclareMathOperator{\TestRecall}{\mbox{\footnotesize TestRecall}}
\DeclareMathOperator{\TestRecallWithFlakes}{\mbox{\footnotesize TestRecallWithFlakes}}
\DeclareMathOperator{\ChangeRecall}{\mbox{\footnotesize ChangeRecall}}
\DeclareMathOperator{\SelectionRate}{\mbox{\footnotesize SelectionRate}}
\DeclareMathOperator{\Score}{\mbox{\footnotesize Score}}
\DeclareMathOperator{\LikelyFailing}{\mbox{\footnotesize LikelyFailing}}
\DeclareMathOperator{\HighlyRanked}{\mbox{\footnotesize HighlyRanked}}
\DeclareMathOperator{\ScoreCutoff}{\mbox{\footnotesize ScoreCutoff}}
\DeclareMathOperator{\CountCutoff}{\mbox{\footnotesize CountCutoff}}

\title{Predictive Test Selection}

\author{
\IEEEauthorblockN{Mateusz Machalica, Alex Samylkin, Meredith Porth, Satish Chandra}
\IEEEauthorblockA{Facebook, Inc.\\
\{stupaq, bane, mporth\}@fb.com, schandra@acm.org}
}

\begin{document}

\maketitle

\begin{abstract}
Change-based testing is a key component of continuous integration at Facebook.
However, a large number of tests coupled with a high rate of changes committed
to our monolithic repository make it infeasible to run all potentially-impacted
tests on each change.
We propose a new \emph{predictive test selection strategy} which selects a
subset of tests to exercise for each change submitted to the continuous
integration system.
The strategy is \emph{learned} from a large dataset of historical test outcomes
using basic machine learning techniques.
Deployed in production, the strategy reduces the total infrastructure cost of
testing code changes by a factor of two, while guaranteeing that over 95\% of
individual test failures and over 99.9\% of faulty changes are still reported
back to  developers.
The method we present here also accounts for the non-determinism of  test
outcomes, also known as test flakiness.
\end{abstract}


\section{Introduction} \label{sec:introduction}

Like many other organizations, Facebook maintains a monolithic repository for
code development.
This means that any code change committed by a developer must first ensure that
all the potentially impacted code continues to build fine and all the
potentially impacted tests continue to pass.
As a rough indication of the magnitude of the problem,
each of several tens of thousands of changes submitted to our mobile codebase
every week potentially impacts of the order of ten thousands tests that would
need to be exercised.
This renders carrying out exhaustive quality control on each code change
impractical.
To reduce the infrastructure costs of testing changes submitted by developers,
as well as to speed up delivery of correctness signal,
change-based test selection techniques are inevitable \cite{Memon2017}.

\begin{figure}[ht]
\includegraphics[width=\linewidth]{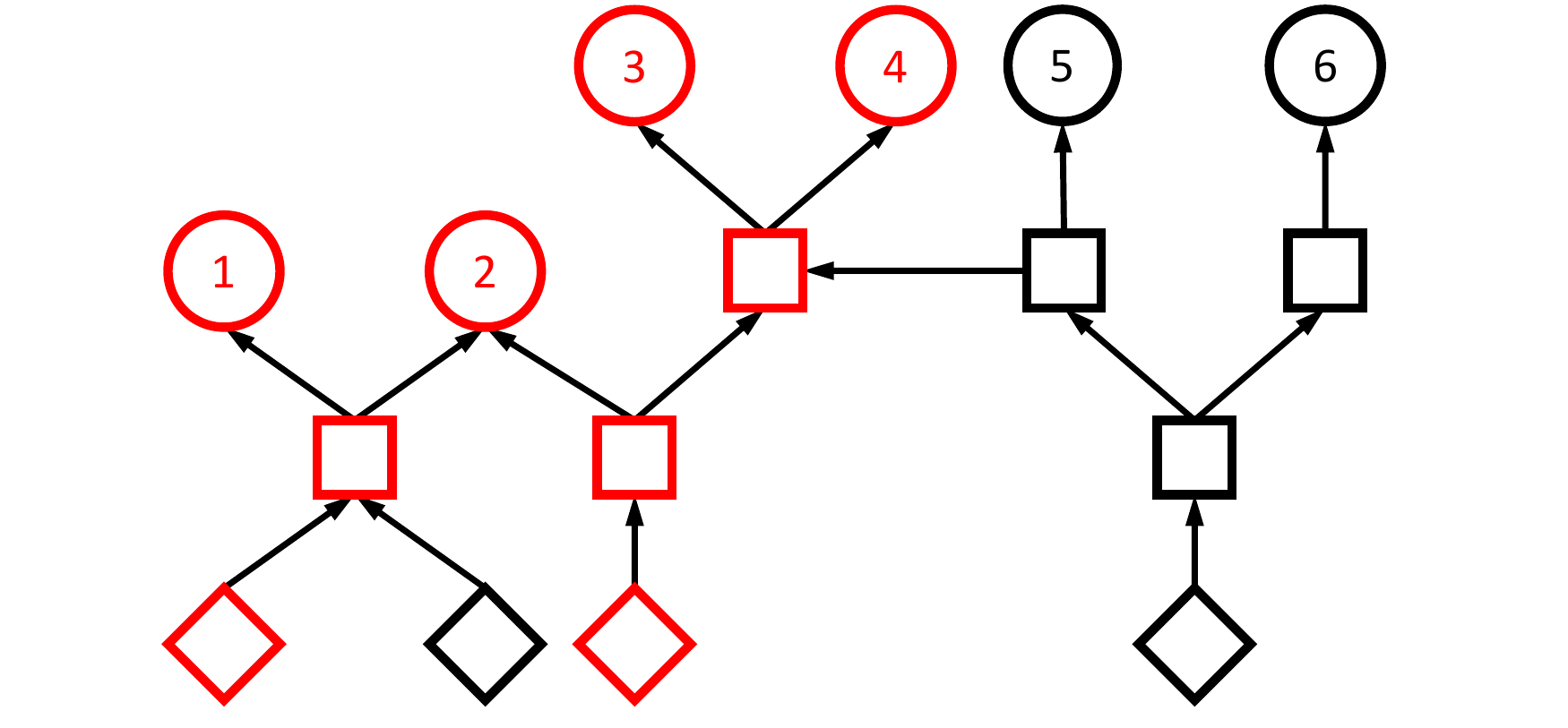}
\caption{
  Example build dependency graph.
  Circles represent tests,
  squares -- intermediate units of code, for example libraries,
  diamonds -- individual source files in the repository.
  An arrow connects two entities $A \rightarrow B$ if and only if $B$ directly
  depends on $A$, which we interpret as $A$ impacting $B$.
  Two files modified in a particular code change and all entities transitively
  dependent upon them are colored red.
  A trivial test selection strategy based on build dependencies would in this
  scenario exercise tests 1, 2, 3 and 4.
  Tests 5 and 6 are not exercised, as they do not depend on modified files.
  \label{fig:build-dependencies1}
}
\end{figure}

A common change-based test selection strategy is to choose tests that
transitively depend on \emph{modified} code according to build dependencies,
as seen in \autoref{fig:build-dependencies1}.
This technique has been employed at Facebook for two main reasons.
First, dependency information recorded in build metadata makes it very easy to
identify all tests transitively depending on modified files.
More importantly, strict build isolation enforced by the build system guarantees
this technique identifies all tests that could possibly be impacted by the
change.
A major disadvantage of this approach is the number of tests selected on code
changes, in the order of $10^4$.
This is despite a vast majority of code changes touching only a few files and
altering low-hundreds in lines of code.

There exists a spectrum of change-impact analysis methods with varying
applicability to the test selection problem.
The approach based on build dependency graph,
as mentioned above, utilizes build metadata.
One can envision a strategy based on \emph{static analysis} of the code,
at various levels of granularity.
A coarse-grain stategy would be to select all tests that transitively refer to
any modified class in Java;
a finer-grain strategy would analyze transitive data and control dependences.
This later approach could reduce the number of tests selected for a particular
code change, due to finer granularity of dependency information it utilizes.
However, sound treatment of certain language features,
such as reflection in Java, may require compromising precision of such
dependency analysis \cite{Legunsen2016}.
Also, this approach is hard to apply in multilingual code bases where it is
possible for program control flow to cross language boundaries,
such as Facebook's mobile repository.

Selecting tests using change impact based on \emph{dynamic analysis} could
potentially further reduce the number of tests exercised on a particular change.
Ekstazi \cite{Gligoric2015} is an example of a test selection approach based on
dynamically collected information about classes loaded into JVM during a
previous execution of a particular test.
Note that all dynamic methods \emph{must} be treated as approximate,
in a sense that they may ignore tests that would detect regression in a
particular change.
This is due to the fact it is fundamentally impossible to know control flow of a
test execution \emph{before} the test is exercised.
Since test selection must take place before the test is run on a particular
version of the code, it cannot be based on dynamic analysis of test's behavior
on \emph{this exact} version.
While it is possible to base test selection on dynamic analysis of the code a
particular change is based on, one can no longer make strict guarantees on
quality of this approach, as even a small code change can arbitrarily alter
runtime behavior.
Also, maintaining per-test code coverage information accurate enough to drive
the test selection process is impractical in large monolithic repositories
\cite{Memon2017}, while recording it requires language-specific infrastructure
and is challenging across language boundaries.

In this work, we propose and evaluate a data-driven,
or \emph{predictive}, test selection strategy.
We look at test selection as a problem of learning a \emph{classifier},
which, given a code change and a test, says how likely is it that the test will
fail on that code change.
Such a classifier is trained based on historical data:
it is learned automatically from previous code changes and test outcomes on
those changes.
Obviously, since the goal is to predict test outcomes for future as-yet-seen
code changes, the classifier acts upon a set of features extracted from code
changes and from tests.
Thus, a future code change's behavior on a test can be predicted based on their
respective features, and the learned behavior of similar code changes or similar
tests.
Such test selection strategy is approximate in the same sense as one based on
dynamic analysis is -- it may fail to select tests that would detect
regressions.
We prefer the predictive test selection over one based on information collected
by dynamic analysis due to unknown accuracy of the latter and challenges
associated with recording runtime information at Facebook's scale.

Predictive test selection requires several important engineering considerations.
Among those, how to account for \emph{flakiness} in test outcomes is a key
hindrance.
Flakiness is the phenomenon whereby the same test produces different outcomes
upon multiple independent trials.
If we train our classifier on flaky test results,
we get very poor predictions.

In this paper, we describe in detail the design of our predictive test
selection system, as well as the way in which we compensate for flakiness.
Our key results are the following:
\begin{itemize}

  \item Using predictive test selection, we can catch over 95\% of individual
  test failures and over 99.9\% of faulty code changes.
  A code change is marked faulty if \emph{any} of the individual tests run in
  response to the code change fails;
  typically it is the recall of the faulty code changes that matters.

  \item The test selection procedure selects fewer than a third of the tests
  that would be selected on the basis of build dependencies;

  \item At the same time, we reduce the total infrastructure cost of
  change-based testing by a factor of two.

\end{itemize}
The predictive test selection has been operational at Facebook for several
months.

The rest of the paper is organized as follows.
In \autoref{sec:continuous-integration} we give background information on
Facebook's repository structure and the continuous integration system.
\autoref{sec:learning-to-select-tests} discusses approximating a set of tests
impacted by a particular change.
In \autoref{sec:test-selection-model} we describe the process of training and
evaluating the predictive test selection strategy.
\autoref{sec:test-flakiness} discusses how test flakiness affects the learned
model.
We present measured performance of the strategy in \autoref{sec:results}.
In \autoref{sec:related-work} we cover related work and conclude by discussing
future directions in \autoref{sec:conclusions-and-future-directions}.

\section{Continuous Integration at Facebook} \label{sec:continuous-integration}

\subsection{Repository Structure}

At Facebook, all of source code comprising mobile applications and backend
services is stored in a single code repository.
There are few restrictions on interdependencies between different portions of
the code base, to the point where Facebook apps for iOS and Android platforms
utilize the same C++ libraries as the company's backend services.
However, any form of versioning parts of the repository is not supported and
strictly disallowed.
So is creating merge commits or maintaining long-lived feature branches.
This forces all code changes submitted by developers to be linearized in a
single \emph{master} branch of the repository,
also known as \emph{trunk-based development} \cite{Potvin2016}.

The monolithic code base model brings a number of benefits \cite{Potvin2016},
it:
simplifies dependency management, and in particular,
makes it impossible for the dreaded diamond dependency problem to occur;
forces end products to migrate to recent versions of libraries they depend on;
enables large-scale refactorings of library APIs and all callsites,
in one atomic change;
encourages engineers to make small, incremental changes as well as check for
interactions between changes made by different developers to the same product
early in the development cycle.
A primary disadvantage of this model is that even small changes can impact a
large number of artifacts, which poses a challenge for scaling developer tools
such as continuous integration.

Within the repository, source code is organized into small and reusable units
referred to as \emph{targets}, each defined in one of many build metadata files
stored in the repository.
Each target describes how to materialize its output artifacts from declared
inputs, such as source files, and other targets it depends on.
At the time materialisation of a particular artifact (e.g.
compiling and linking a test binary) is requested,
our cross-platform, language-agnostic build system,
reads all relevant build metadata files in the repository and executes recipes
prescribed in target specifications until the requested artifact is
materialized.
Targets are materialized in an order obtained by topologically sorting a
directed acyclic graph (DAG) of dependencies between targets,
with independent targets being materialized in parallel when possible.

Automated tests defined by developers are organized in \emph{test targets}
similarly to code under test.
Those special targets define both how to materialize executables implementing
tests as well as how to exercise them.
When a developer or continuous integration system requests certain test to be
run, the build system takes care of the whole process by first building the
binaries and then launching them in an appropriate environment,
for example using a mobile device emulator.

Depending on the programming language and testing framework in question,
a single test target may define multiple \emph{test cases},
each making a number of \emph{assertions}.
In our presented work we have considered a test target be the atomic unit of
verification performed in the continuous integration system.
We consider a test target to \emph{fail} if and only if any assertion made by
any of test case has failed or the test binary terminated prematurely,
for example due to a process crash.

\subsection{Testing in Developer Workflow}
\label{sec:testing-in-developer-workflow}

Typical mobile developer workflow involves:
\begin{enumerate}

  \item Creating a \emph{change} based on a recent commit in the master branch.
  Each change embeds information on the version of the repository it is based
  on.
  It is thus possible to reconstruct exactly the state of the code base after
  the change.

  \item Creating a \emph{diff} in the internal code review tool,
  attaching the change as the first \emph{version} of the diff.

  \item Iterating on the diff based on review feedback,
  creating a new version on each iteration.

  \item Once the diff is accepted, submitting it for \emph{landing}, which
  involves asynchronously pushing the diff into the master branch, if it
  introduces no detectable breakages.

  \item If the diff is rejected during land,
  for example due to rebase conflicts or errors detected by static or dynamic
  analysis, the author of the diff may continue iterating or abandon it.

\end{enumerate}

Automated testing happens at all stages of the developer workflow,
with objectives varying from stage to stage.

\subsubsection{Pre-submit}

Although left with freedom to skip this step,
developers typically exercise a few hand-picked tests prior to creating a diff
in the code review tool.
In this way engaging reviewers is avoided if the diff is broken in a quickly
detectable way, which limits human resources used in the review process.

\subsubsection{Diff-time}

Facebook's continuous integration system automatically runs a subset of tests
every time a new version of a diff is created and reports their results in the
code review tool.
Each tested patch is first rebased onto a recent version of master branch that
is known to pass all automated tests, which guarantees reported test failures to
indicate diff introducing a regression.
The developer need not wait for the results before working on a follow up diff.
This feedback lets developers fix any detected problems before they move on to a
different task and lose some of the context of the change.
Ideally results of test suite should be delivered in no more than ten minutes.

\subsubsection{Land-time} \label{sec:land-time}

Once a diff is submitted for landing, it is rebased onto a recent version of
master branch that passes all automated tests and a (possibly more
comprehensive) subset of tests is run on the modified version of the code base.
The diff is rejected if any of the tests reports a failure on it.
This stage of testing acts as a gatekeeper, guarding against breakages slipping
into the master branch.

Note that due to the velocity of code changes,
it is not feasible to serialize the process of land-time testing for all of
them.
This implies that a number of code changes submitted for landing will be rebased
onto the same version of master branch, tested in parallel and then serialized
into a linear history of commits.
It is possible that changes being landed simultaneously pass when tested
individually, even though they would cause test failures when rebased on top of
each other.

\subsubsection{Stabilization}

Once every few hours, all tests are exercised on the most recent version of
master branch.
Tickets are created for failing tests, which are then triaged either to their
respective owners or authors of breaking changes.
This stage aims to catch any breakages that slipped through prior stages and
find versions of master branch that are free of bugs detectable using automated
tests.
Release candidates of mobile applications can only be based of such versions of
the repository, which implies no bug detectable via automated testing can affect
the quality of released product, even if it slips through prior stages.

The stabilization stage could also be considered a form of testing all diffs
submitted in the past few hours \emph{in a batch}.
Testing multiple diffs at once can greatly reduce infrastructure cost of
continuous integration, although it does have a few notable disadvantages.
Successful completion of a test suite does not imply each diff in the sequence
is free of detectable faults, as it is possible for the sequence to contain a
diff introducing a breakage and a following diff fixing the bug.
Additionally, it the test suite detects a fault it is usually not immediately
clear which of the diffs was a culprit \cite{Micco2017}.

Out of all discussed stages of automated testing,
diff- and land-time ones require an order of magnitude more machine resources
than testing during the stabilization stage.
Resource requirements of earlier stages scale at least linearly in the number of
developers, contrary to resources devoted to the last stage.
This is due to the relative frequency of events that trigger different stages,
which increases for those happening earlier in the developer workflow.

Presence of the stabilization stage means the main goal of diff- and land-time
testing is to boost developer productivity at an additional infrastructure cost,
rather than to reduce risk of buggy software being released.
Continuous integration system must work with a trade-off between infrastructure
cost and latency of test signal, as well as chances of a breaking change being
landed, both of which should be minimized.
Thoroughness of testing at each stage can be controlled to decide which part of
the trade-off is implemented.
\begin{itemize}

  \item Less thorough testing at diff- and land-time would cause developers to
  learn about errors that need to be corrected at a later time,
  in the extreme case at the stabilization phase.
  This increases the need to context-switch between tasks, which negatively
  impacts developer productivity.

  \item More thorough testing prior to landing a diff,
  although reducing the chances of detectable bug being committed to master
  branch, negatively impacts the cost of testing and/or latency of correctness
  signal provided to a developer.

\end{itemize}

\section{Learning to Select Tests} \label{sec:learning-to-select-tests}

\subsection{Approximating The Set of Impacted Tests}

Implementing a \emph{perfect} test selection strategy,
that is one choosing all-and-only tests impacted by a particular code change,
is not feasible.
Such a strategy would necessarily require access to pieces of information
unavailable at test selection time.
Our key observation is that we can \emph{derive from data} a strategy that is
close enough to a perfect one.

\emph{While we cannot compute exactly the set of impacted tests for a particular
change, we can approximate this computation by learning to identify which tests
would have reported a failure, based on historical data.}

It is important to note that our approximation needs not be \emph{conservative},
in a sense that it may miss some of the impacted tests,
yet it can still be applied in diff- and land-time testing.
As explained in \autoref{sec:continuous-integration}, these stages of testing
are not only non-essential, but also cannot maintain the quality of the released
product on their own due to races between diffs being tested and landed
simultaneously.
Their primary objective is to improve developer productivity at some additional
infrastructure cost.
This means the continuous integration system \emph{may} trade how thorough
testing is applied at each stage with how much it costs in terms of machine
resources and developer time, in order to optimize developer experience while
keeping these costs in check.

\begin{figure}[ht]
\includegraphics[width=\linewidth]{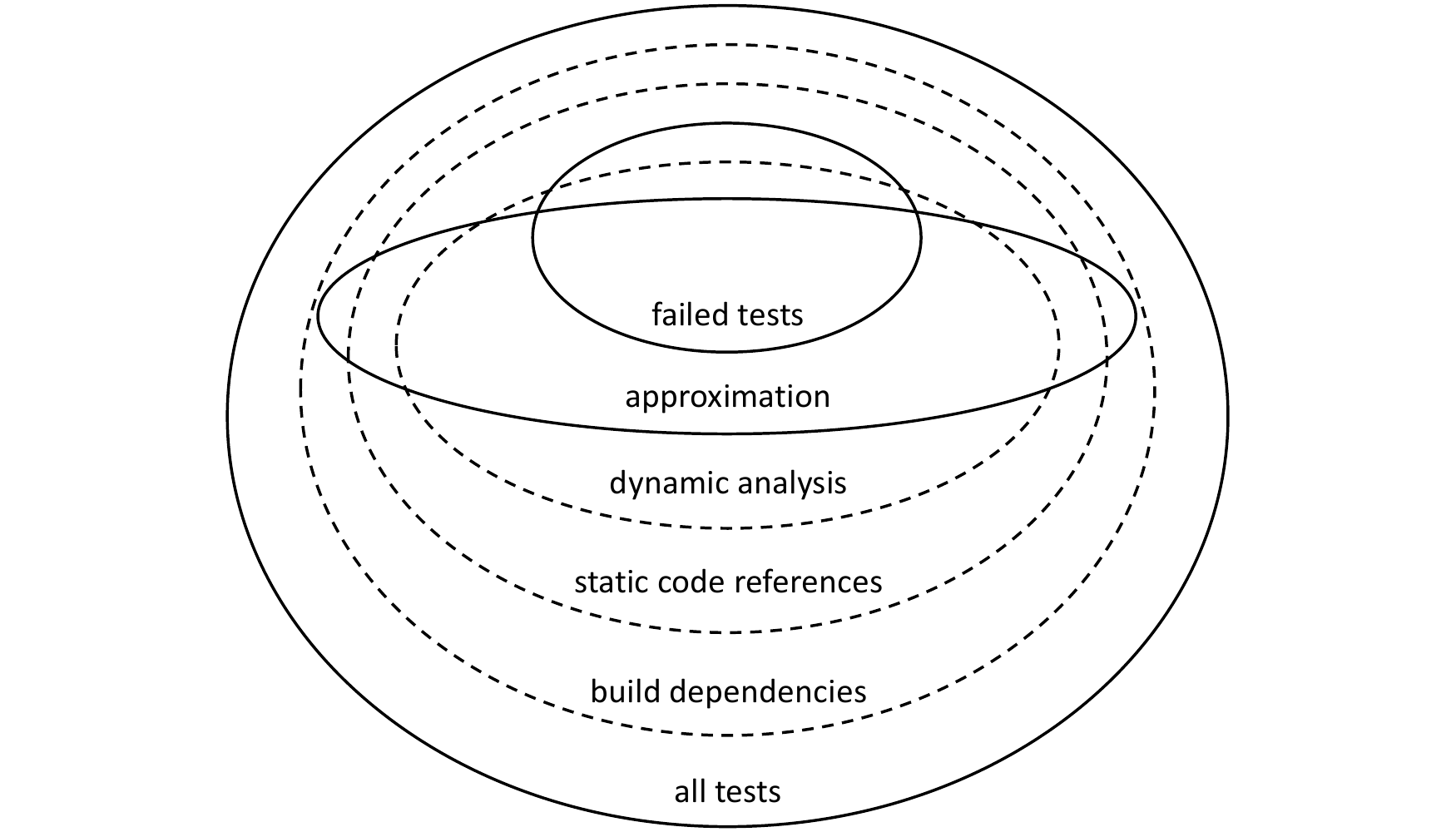}
\caption{
  Schematic relationship between sets of tests selected by different methods for
  a hypothetical change, as well as all and only failed tests.
  Proportions are not preserved.
  Note that approximate test selection is not conservative,
  in a sense that it has missed a portion of failing tests,
  which definitely were impacted by the change.
  Similarly, strategy based on dynamic analysis has missed a portion of
  failures, as it has selected a subset of tests based on information collected
  on prior version of the code.
  Coarse-grained build dependencies, specified at the granularity of targets,
  rather than individual files, typically cause test selection based on analysis
  of code references more accurate than one based on build dependency graph.
  \label{fig:impacted-outcomes}
}
\end{figure}

\autoref{fig:impacted-outcomes} depicts test targets impacted by a code change
according to various test selection strategies as well as their respective
outcomes.
While all tests reporting failures must be impacted by the change,
not all of the tests that passed are.
Intuitively, when learning to approximate the set of impacted tests for a
particular change, we should aim to capture as many tests that would report
failure and as few unimpacted tests as possible.
We will now formalize this intuition and discuss metrics that quantify
usefulness of test selection strategy to the process of change-based testing in
continuous integration system.

\subsection{Measuring Quality of Test Selection}

At diff- and land-time stages of testing, the continuous integration system does
not need to exercise all test targets transitively depending on code modified in
a diff.
While the fact that specific tests are passing may constitute useful feedback to
developers, it suffices if only targets that \emph{would} fail are run on a
particular code change.
This is due to the code change being ineligible for landing and requiring
developer action \emph{if and only if} it breaks any test.
For the same reason, had we known ahead of time that none of the tests would
report a failure, we would not need to run any of them.
Since almost 99.9\% of test targets selected by build-dependency-based selection
strategy pass, selecting fewer passing tests could greatly reduce the resources
consumed by testing.

We can formalize the above considerations by defining three metrics which
quantify quality of a particular test selection strategy.
Let us introduce notation we will use throughout the rest of this paper.
For a test selection strategy $s$ and a code change $d$,
let:
\begin{itemize}

  \item $\AllTests(d)$ be the set of test targets present in the version of the
  repository associated with $d$,

  \item $\DependentTests(d) \subseteq \AllTests(d)$ be the set of test targets
  transitively dependent upon any file modified in $d$ according to build
  metadata,

  \item $\SelectedTests(s, d) \subseteq \AllTests(d)$ be the set of test targets
  selected by $s$ on $d$,

  \item $\FailedTests(d) \subseteq \DependentTests(d)$ be the set of test
  targets that would report failure on $d$ had all tests been exercised,

\end{itemize}

\begin{definition}[Test recall]

  Let $s$ be a test selection strategy and $D$ a set of code changes,
  such that for $F_d = \FailedTests(d)$,
  $\exists_{d \in D} F_d \neq \emptyset$.
  \begin{equation*}
    \TestRecall(s, D) = \frac{
      \sum_{d \in D} {|\SelectedTests(s, d) \cap F_d|}
    }{
      \sum_{d \in D} {|F_d|}
    }
  \end{equation*}

\end{definition}

Intuitively, test recall equals empirical probability of a particular test
selection strategy ``catching'' an individual failure.

\begin{definition}[Change recall]

  Let $s$ be a test selection strategy and $D$ a set of code changes,
  such that for $F_d = \FailedTests(d)$,
  $\exists_{d \in D} F_d \neq \emptyset$.
  \begin{equation*}
    \ChangeRecall(s, D) = \frac{
      |\{ d \in D \mid \SelectedTests(s, d) \cap F_d \neq \emptyset \}|
    }{
      |\{ d \in D \mid F_d \neq \emptyset \}|
    }
  \end{equation*}

\end{definition}

Intuitively, change recall equals empirical probability of a particular test
selection strategy ``catching'' at least one failure on a faulty code change.

\begin{definition}[Selection rate]

  Let $s$ be a test selection strategy and $D$ a set of code changes.
  \begin{equation*}
    \SelectionRate(s, D) = \frac{
      \sum_{d \in D} {|\SelectedTests(s, d)|}
    }{
      \sum_{d \in D} {|\DependentTests(d)|}
    }
  \end{equation*}

\end{definition}

Note that selection rate measures the fraction of test targets selected by a
particular strategy relative to the build-dependency-based one.
Since the latter is easily computable and identifies all (but not only) the
impacted test targets for each code change, it constitutes a good baseline to
compare other methods against.

Whenever a particular set of changes $D$ follows from the context,
we omit it and write $\TestRecall(s)$, $\ChangeRecall(s)$,
$\SelectionRate(s)$ respectively.

\section{Test Selection Model} \label{sec:test-selection-model}

In this section, we present our main contribution:
a statistical model that selects a subset of tests to exercise on a particular
code change.
The model, rather than being defined manually,
is derived using basic machine learning techniques from a large dataset that
records the outcomes of running all potentially impacted tests on a sample of
code changes submitted to the continuous integration system.

Ideally, we would like to learn a model that selects all-and-only-those tests
impacted by a particular code change.
As discussed in \autoref{sec:learning-to-select-tests} we cannot observe exactly
which tests were impacted by the change even if we exercise all of them.
We can, however, tell which tests have failed on the change.
Since all failed tests must have been impacted,
we can instead learn to predict whether a particular test would have failed and
select tests that have high likelihood of failing according our prediction.

Having access to large dataset containing outcomes of tests run on historical
changes submitted to the continuous integration system,
we can train a binary classifier, which recognizes pairs of code changes and
tests that reported failure on them.
While doing so, we must ensure that the classifier generalizes to previously
unseen changes, as it is extremely unlikely for the same historical code change
based on the same version of the repository to ever be created again.
Therefore, we make the classifier operate on a set of features in place of an
actual code change and an actual test.
The trained classifier is a function that takes as inputs features of a given
change and a test, and outputs a likelihood of the actual test failing on the
change if it was run.

\begin{figure}[ht]
\includegraphics[width=\linewidth]{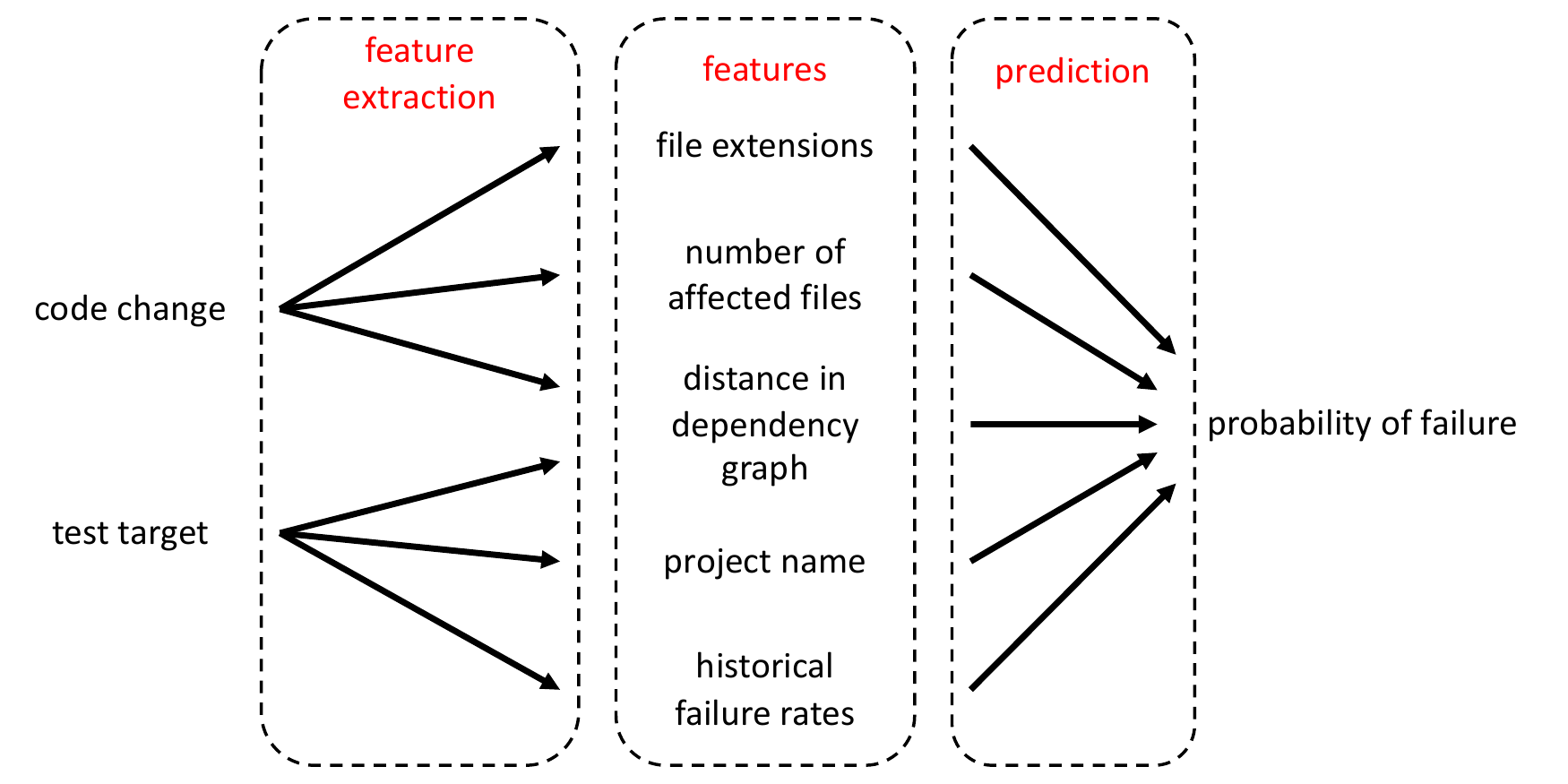}
\caption{
  A schematic explanation of feature extraction and prediction process.
  Individual features used by the test selection model are covered in
  \autoref{sec:feature-engineering}.
  \label{fig:feature-extraction}
}
\end{figure}

We are using the learned test selection to make data-guided trade-off between
the cost and quality of test signal at diff- and land-time stages of testing in
the continuous integration system.
For this reason, we must be able to predict behavior of the test selection model
on changes submitted in the future and adjust it,
in order to achieve desired correctness guarantees.

\subsection{Feature Engineering} \label{sec:feature-engineering}

Model inputs, a change $d$ and a test target $t$ provide a natural way to think
about different categories of features:
change- and target-dependent ones, as well as cross features between them.

Change level features consist of:
\begin{itemize}

  \item \emph{Change history for files} is useful to identify active areas of
  development which are more prone to breakages.
  We thus use features indicating number of changes made to modified files in
  the last 3, 14, and 56 days.

  \item \emph{File cardinality}, or number of files touched in a change.
  Large changes are harder to review and we assume that probability of a test
  failure is lower for small changes.

  \item \emph{Target cardinality}, i.e.
  number of test targets triggered by a change.
  If certain files are used in many projects then a small change in them might
  trigger unexpected behavior.

  \item Our projects use multiple programming languages,
  which have different breakage patterns.
  We use a fixed-size bit vector to identify \emph{extensions of files} modified
  in a change.

  \item Number of \emph{distinct authors} for files in a change might indicate
  common code that is used in multiple project and requires extra attention.

\end{itemize}

Target level features consist of:
\begin{itemize}

  \item \emph{Historical failure rates} of a target are a good baseline for the
  probability of failure.
  We include a vector of failure rates in the last 7,
  14, 28 and 56 days as a feature.

  \item \emph{Project name} is useful to identify an area the target covers and
  categorize breakage patterns based on a project.

  \item \emph{Number of tests} in a target can be used as a proxy of the code
  area covered by it.

\end{itemize}

Cross features are:
\begin{itemize}

  \item \emph{Minimal distance} between one of the files touched in a change and
  the prediction target.
  The feature approximates how close are changes to a given target and the
  significance of the impact on it.

  \item Number of \emph{common tokens} shared by paths of modified files and
  test defines lexical distance to proxy human perceived relevance.

\end{itemize}

\subsection{Model Architecture} \label{sec:model-architecture}

Our learned test selection strategy is based on gradient boosted decision trees
classifier \cite{Chen2016}.
This learning algorithm has a number of properties desirable for our use-case:
it does not require normalizing feature values,
takes little time to train on available hardware,
works out-of-the-box for datasets where numbers of positive and negative
examples differ by a few orders of magnitude, supports ordinal and categorical
features.

The classifier is learned on test outcomes recorded for changes submitted over
past three months.
Each entry in such \emph{training dataset} represents a change $d$ and a test
target $t \in \DependentTests(d)$, and is labeled as positive if and only if $t
\in \FailedTests(d)$.
The classifier provided with features extracted from a particular code change
$d$ and test target $t$ returns a score $\Score(d, t) \in [0, 1]$,
which can be interpreted as estimated likelihood of $t \in \FailedTests(d)$.

The proposed strategy $s^*$ constructs a subset of selected targets for a
particular change $d$ based on scores returned by the classifier for all $t \in
\DependentTests(d)$.
It is parameterized with a score threshold above which the target shall be
selected, $\ScoreCutoff(s^*) \in [0, 1]$,
and a number of top-scoring targets to select for each change,
$\CountCutoff(s^*) \in \mathbb{N}_{\ge 0}$.
\begin{itemize}

  \item $\LikelyFailing(s^*, d)$ contains all $t \in \DependentTests(d)$ for
  which $\Score(d, t) \ge \ScoreCutoff(s^*)$.

  \item $\HighlyRanked(s^*, d)$ contains up to $\CountCutoff(s^*)$ of
  $t \in \DependentTests(d)$ with highest $\ScoreCutoff(s^*)$.

\end{itemize}
The final strategy $s^*$ is defined as a union of the two approaches above
$\SelectedTests(s^*, d) = \LikelyFailing(s^*, d) \cup \HighlyRanked(s^*, d)$.

\subsection{Model Calibration} \label{sec:model-calibration}

Behavior of trained classifier depends on training dataset and chosen learning
algorithm.
Feature engineering, collecting higher quality and quantity of data,
tunning hyper-parameters of the learning algorithm all contribute to the
classifier returning more accurate scores.
The more accurate the scores, the better is the trade-off between correctness
and cost savings brought by predictive test selection.
However, actual performance of proposed strategy $s^*$ is determined by values
of $\ScoreCutoff(s^*)$ and $\CountCutoff(s^*)$ chosen during \emph{calibration}
based on the desired performance.

We use strategy $s^*$ trained on past code changes to select tests for changes
created in the future.
Therefore, when calibrating $s^*$ and evaluating its performance we must use
test results not included in the training dataset.
We split collected data, such that test outcomes recorded for changes submitted
during the most recent week fall into the \emph{testing dataset} and the
remainder forms the \emph{training dataset}.
Described approach ensures the evaluation procedure closely replicates how the
model is going to be used in practice, which makes estimated model performance
match closely the one observed in production.

Accurately measuring test and change recall for a set of all code changes,
$D$, recently submitted to continuous integration system,
and a particular test selection strategy $s$ requires knowing outcomes of all
test targets belonging to $\DependentTests(d)$ for each $d \in D$ in order to
determine $\FailedTests(d)$.
Note that had we only exercised $\SelectedTests(d)$ we would not be eable to
determine whether we had missed any test targets belonging to $\FailedTests(d)
\setminus \SelectedTests(d)$.
Thus, the only way to calculate test and change recall of $s$ is to exercise all
test targets in $\DependentTests(d)$ for each change $d \in D$.
The need to repeatedly evaluate performance of the test selection model renders
this approach impractical.
Also, running all possibly impacted tests on each change defeats the purpose of
our work, which is to significantly reduce the infrastructure cost of continuous
integration.
In practice, we have found it is sufficient to estimate the performance of a
test selection strategy based on a sample of test results.
We sample independently a subset $D' \subset D$ such that $|D'| \ll |D|$ and
schedule for each $d \in D'$ a \emph{learning test run}.
During such run we exercise all test targets in $\DependentTests(d)$ and record
their results.
We then compute test and change recalls, as well as selection rate of $s$ on
changes in $D'$ and assume they are a good approximation of performance of $s$
on $D$.

Learning test runs do not produce any output visible to developers interacting
with the continuous integration system.
This gives us an opportunity to defer them to off-peak hours,
when load on CI and other developer tools drops significantly.
In this way are able to sample close to a quarter of submitted code changes and
collect evaluation data without increasing peak resource usage of the system,
utilizing only off-peak, spare capacity.

\subsection{Deployment Process} \label{sec:model-deployment}

Since we evaluate the model's performance on only one week worth of test
outcomes, we cannot guarantee it remains unchanged for a much longer period of
time.
For this reason, we have automated the following process to occur on a weekly
basis:
\begin{itemize}

  \item Train new model as described in \autoref{sec:model-calibration},
  including freshly collected data.

  \item Assert that the trained model's performance meets predefined criteria.
  We may require $\SelectionRate(s^*) < 0.3$ for $\TestRecall(s^*) = 0.9$ and
  $\CountCutoff(s^*) = 0$, for example.

  \item In the case that the assertion is violated,
  the responsible team member is notified to investigate regression.

  \item The model meeting the criteria automatically replaces one operating in
  production.

\end{itemize}

The ability to retrain a test selection strategy is a major advantage of a
learning-based approach over a manually devised heuristic,
as the former can dynamically \emph{adapt} to evolving code base and
continuously guarantee a predefined level of correctness.
Automating the process of training, verifying and deploying the model has
reduced the maintenance cost of the system and the likelihood that a human will
cause a model to underperform and affect developer productivity.

\section{Test Flakiness} \label{sec:test-flakiness}

\begin{figure}[ht]
\includegraphics[width=\linewidth]{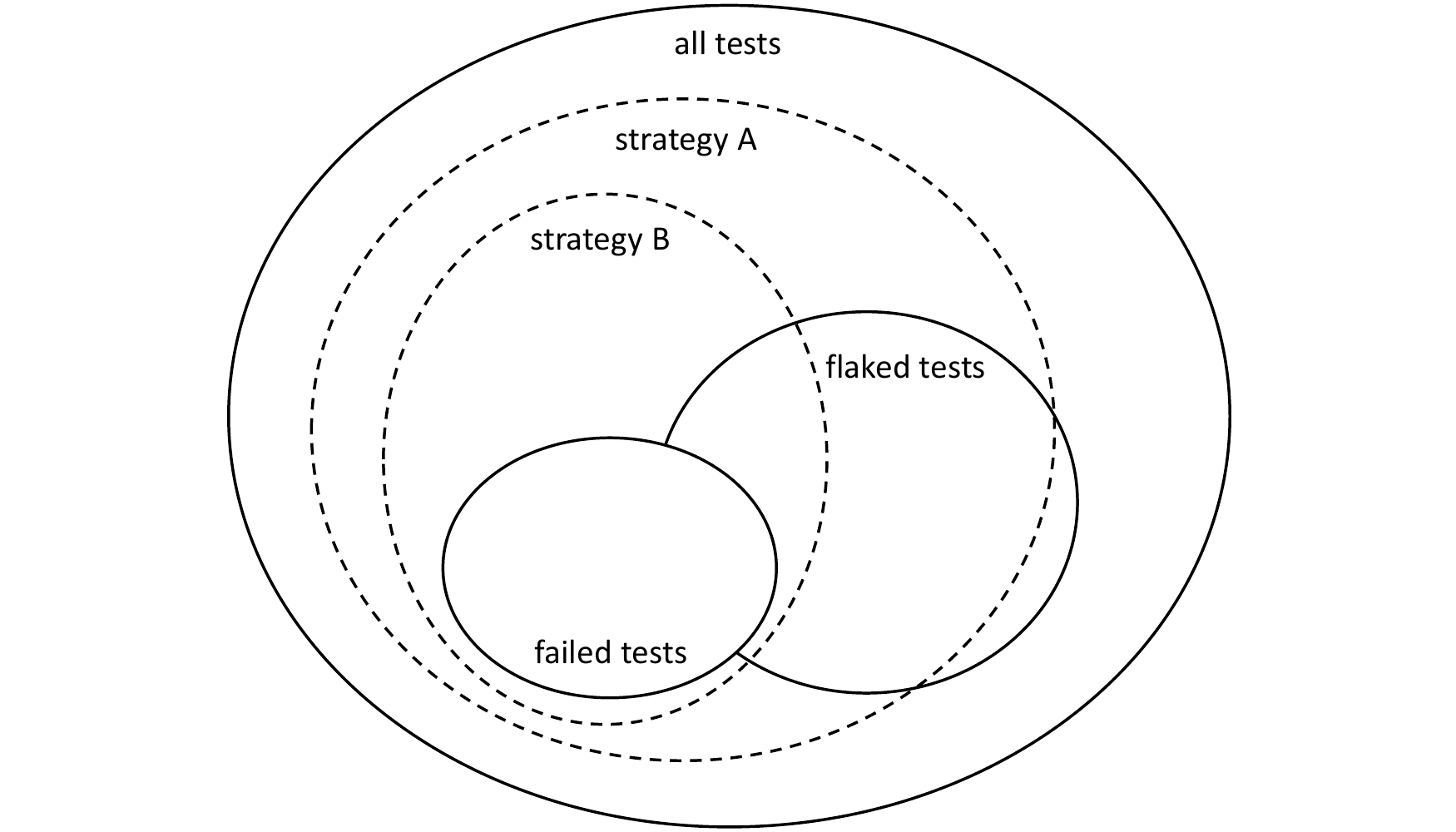}
\caption{
  Schematic relationship between sets of tests selected by two strategies for a
  hypothetical change, as well as tests that failed all and some-but-not-all
  attempts.
  Proportions are not preserved.
  Both strategies capture all failed tests, however the portions of tests that
  failed flakily differ.
  Strategy $A$ having selected more tests overall than $B$ also captured more
  tests failing for reasons unrelated to the change.
  \label{fig:impacted-outcomes-flakiness}
}
\vspace{\baselineskip}
\includegraphics[width=\linewidth]{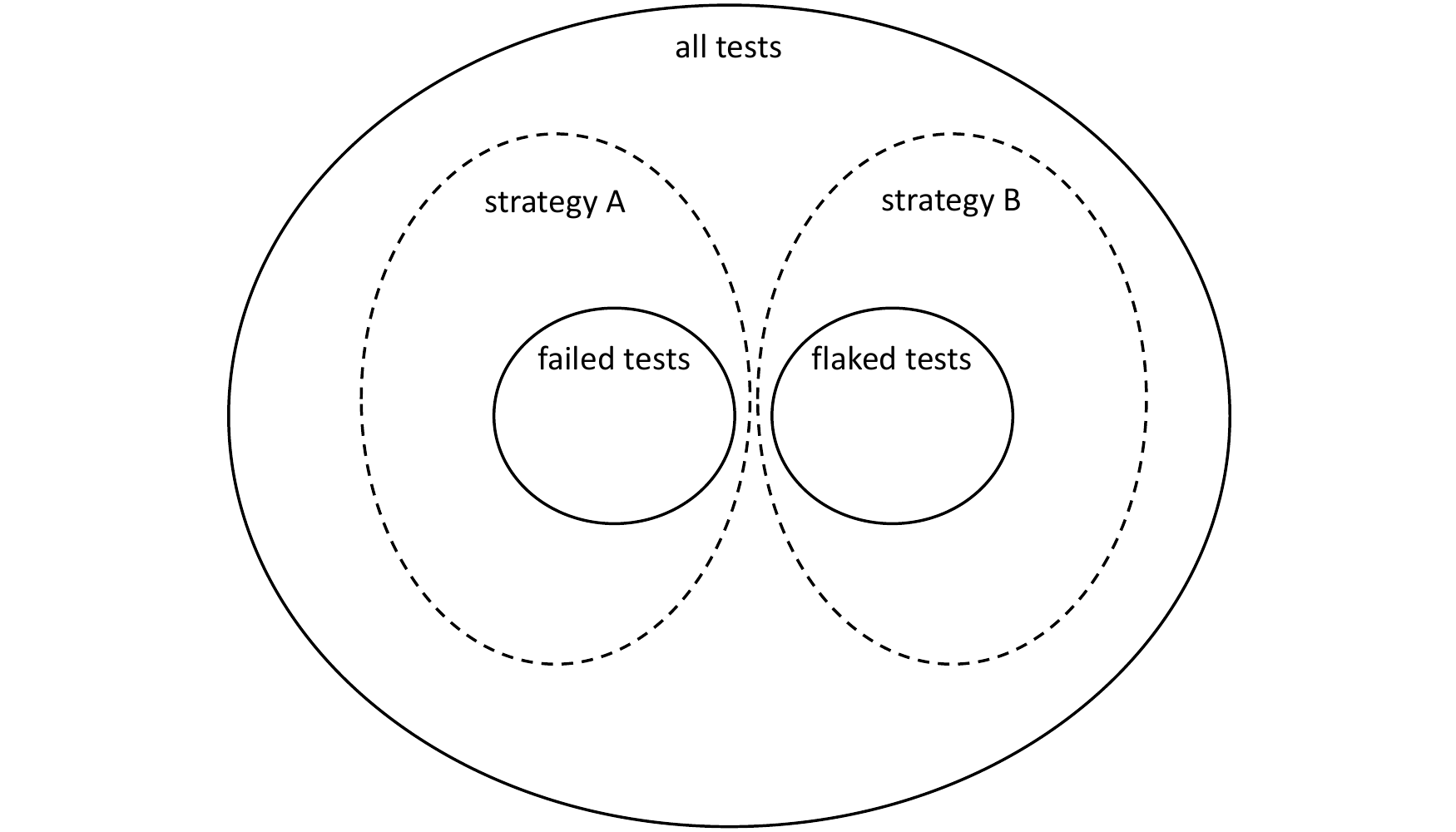}
\caption{
  Schematic relationship between sets of tests selected by two strategies for a
  hypothetical change, as well as tests that failed all and some-but-not-all
  attempts.
  Both strategies captured half of $\FailedTests(d) \cup \FlakedTests(d)$ on a
  change $d$.
  Strategy $A$ captured exactly $\FailedTests(d)$, which is desired behavior,
  while strategy $B$ captured exactly $\FlakedTests(d)$ and none of
  $\FailedTests(d)$, which is not acceptable.
  Note that if we were not able to distinguish $\FailedTests(d)$ from
  $\FlakedTests(d)$, we would measure equal performances of both strategies,
  and thus were not able to deterministically avoid choosing the bad one,
  that is strategy $B$.
  \label{fig:capturing-failures-flakes}
}
\end{figure}

While it's convenient to consider every reported test failure to indicate
presence of a fault, outcomes of real-world tests are frequently affected by
\emph{flakiness}.
In the context of change-based testing, we consider a failure that is not caused
by the change as one caused by test flakiness.
Typical sources of flakiness include \cite{Luo2014, Bell2018}:
usage of random number generators, assumptions about timeliness of asynchronous
operations, races in the test code, reliance on production services,
and tests poisoning the environment.
Although Facebook's developers are incentivized and provided with resources to
write reliable tests and fix flaky ones, we do not believe eradicating test
flakiness entirely is economically viable.
Thus it is a responsibility of the continuous integration system to operate well
in presence of flakiness \cite{Micco2017}.

For the purpose of evaluating test selection strategies,
we identify failures unrelated to the code change by retrying corresponding test
a number of times.
During test runs producing training and evaluation data,
every failed test target is exercised up to ten times or until it reports a
successful result, whichever comes first.
Results of all attempts of each test target $t \in \DependentTests(d)$ are then
aggregated for a particular code change $d$,
so that:
\begin{itemize}

  \item $t \in \FailedTests(d)$ if and only if all attempts failed,

  \item $t \in \FlakedTests(d)$ if and only if there was both failed and
  successful attempt.

\end{itemize}

The described de-flaking procedure assumes that if there exists a possible
execution of a test on a particular version of the code that does not trigger a
failure, then any failure the test may report on that version of the code is
flaky.
This technique is accepted across the industry \cite{Micco2017}.

Let $D'$ be the set of changes sampled for learning test runs.
We have observed that $\sum_{d \in D'} |\FlakedTests(d)|$ is about four times
larger than $\sum_{d \in D'} |\FailedTests(d)|$.
Note that the number of test targets failing flakily depends on the employed
test selection strategy.
For each change $d$, $\FailedTests(d)$ can only contain impacted tests and is
empty for all non-faulty $d$, At the same time,
$\FlakedTests(d)$ may be non-empty irrespective of whether $d$ is faulty or not.
This is due to flaky tests having non-zero chance of reporting failure even if
they are not impacted by a change.
We thus can expect the fraction of failures identified as flaky to increase
significantly if all tests are run on each code change,
and shrink accordingly if a more sophisticated test selection strategy is
employed.
\autoref{fig:impacted-outcomes-flakiness} explains above reasoning on an
example.

It is worth noting that the retry-based de-flaking mechanism may not address all
forms of flakiness.
Hence our estimate on the number of test failures that are not related to the
underlying code change should be treated as lower-bound.

Test flakiness impacts predictive test selection,
as it affects recorded test outcomes used to train and evaluate test selection
models.
In the end of the day, the correctness of test selection strategy depends on its
ability to capture test failures caused by the change,
not flaky ones.
If we were not able to distinguish between them,
we would risk training the model to accurately capture tests that failed
flakily, rather than those that failed detecting a fault.
It is best seen on \autoref{fig:capturing-failures-flakes}.

\section{Results} \label{sec:results}

While performance of the test selection model operating in production is
important for developer productivity and continuous integration resources,
true insight comes from studying how different features affect the trade-off
between the model's ability to catch test failures and the number of tests
needed to be exercised.
In the following sections we use popular model introspection techniques to
determine the impact of utilized features.
Additionally, we cover the impact of test flakiness and show that the learned
test selection strategy performs well in real-world conditions,
where test outcomes are not fully deterministic.

\subsection{Empirical Performance}

\begin{figure}[ht]
\includegraphics[clip,trim=0 0.25cm 0 1cm,width=\linewidth]{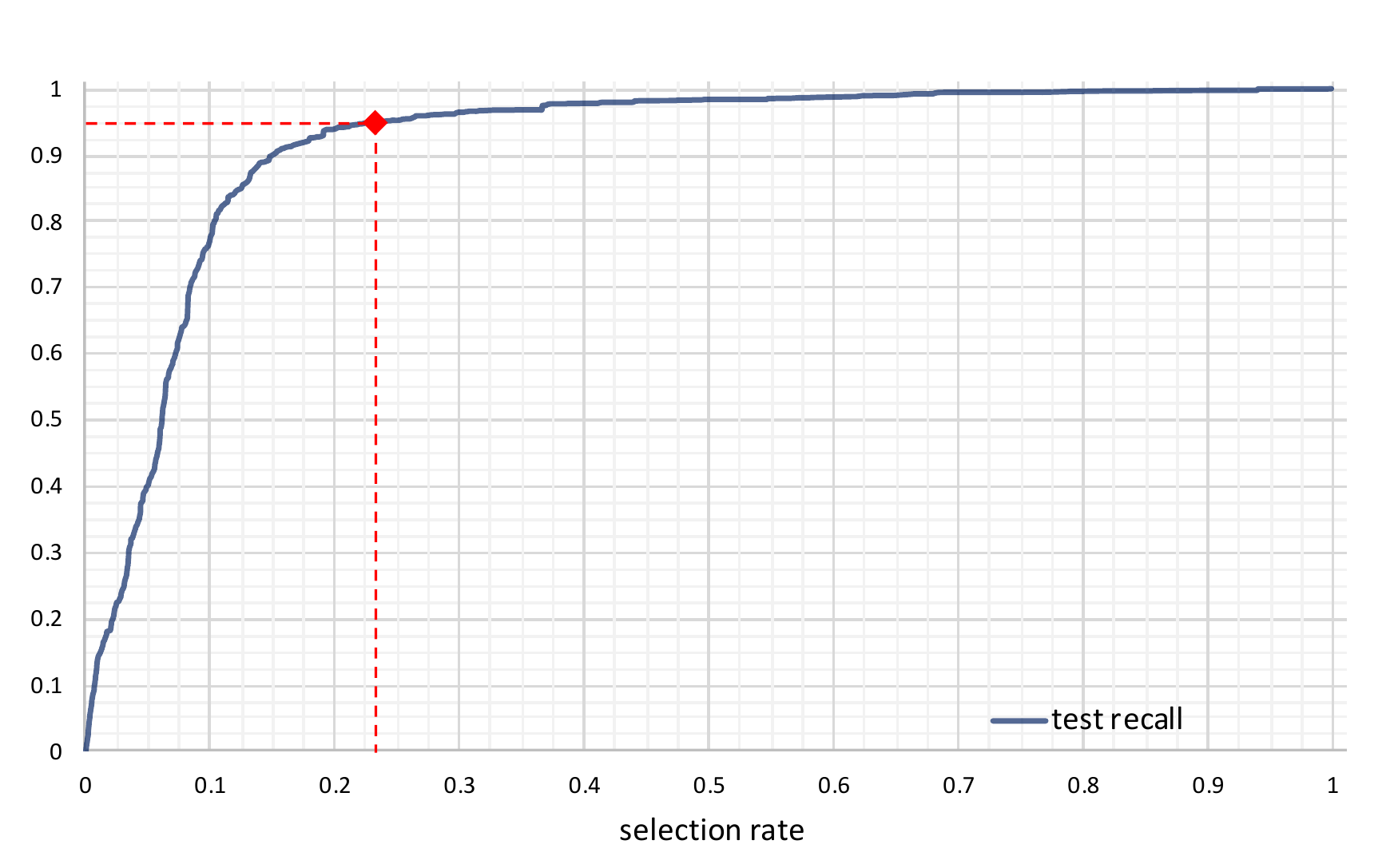}
\caption{
  $\TestRecall(s^*)$, $\SelectionRate(s^*)$ as functions of
  $\ScoreCutoff(s^*)$, whose values vary along the curve.
  We fix $\CountCutoff(s^*) = 0$.
  All relationships can be inverted, so that it's possible to determine
  $\ScoreCutoff(s^*)$ corresponding to a particular value of $\TestRecall(s^*)$
  or $\SelectionRate(s^*)$.
  \label{fig:baseline-score-cutoff}
}
\vspace{\baselineskip}
\includegraphics[clip,trim=0 0.25cm 0 1cm,width=\linewidth]{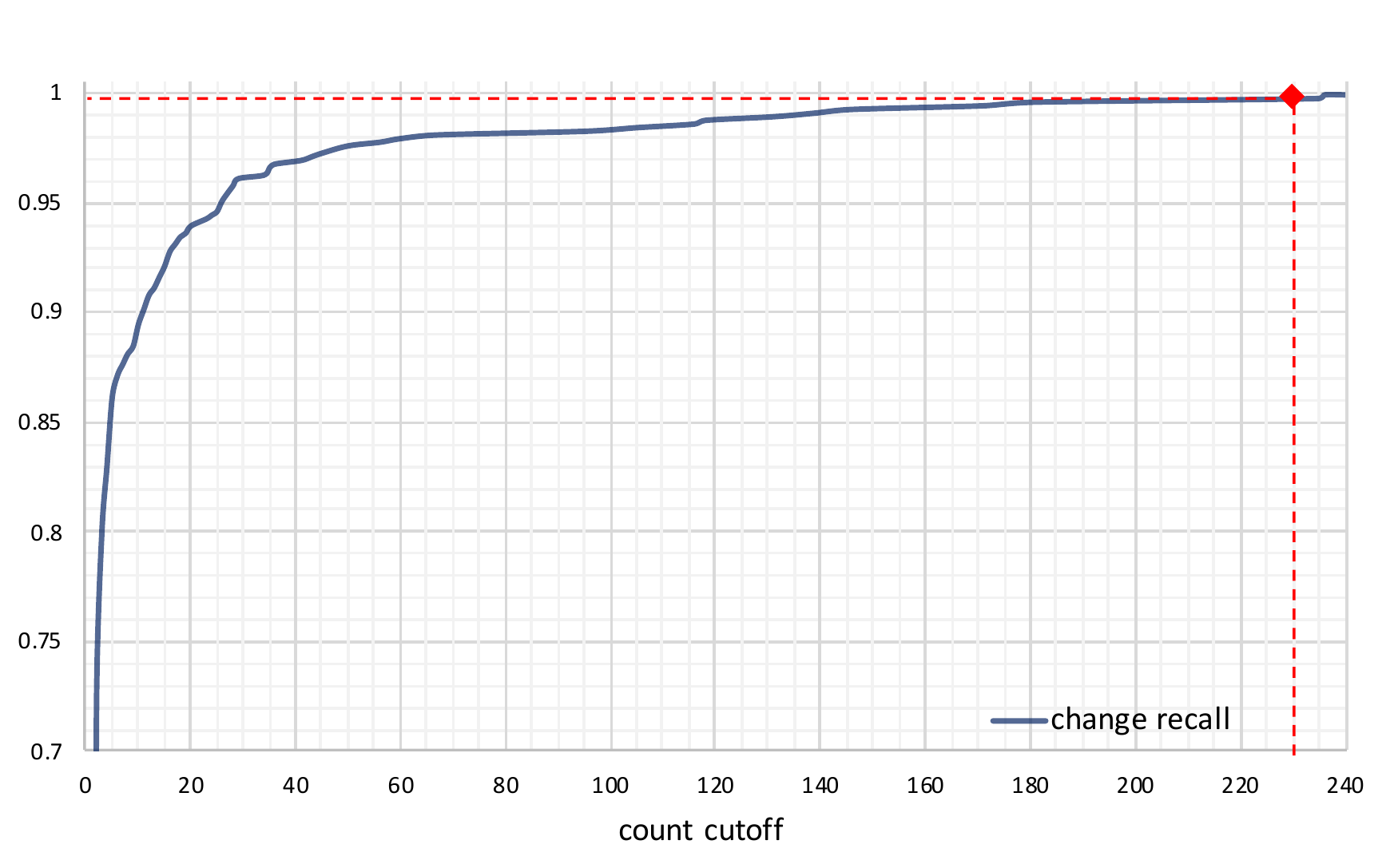}
\caption{
  $\ChangeRecall(s^*)$ as a function of $\CountCutoff(s^*)$,
  for fixed $\ScoreCutoff(s^*) = 0$.
  Note the range on the vertical axis extends from 0.7 to 1.
  This is due to the model being able to capture at least one failure on 70\% of
  faulty changes by selecting only two highest-scoring test targets.
  \label{fig:baseline-count-cutoff}
}
\end{figure}

\autoref{fig:baseline-score-cutoff} and \autoref{fig:baseline-count-cutoff}
depict measured performance of predictive test selection strategy $s^*$ for
varying values of parameters.
We measure dependency of $\TestRecall(s^*)$ on $\ScoreCutoff(s^*)$ and
$\ChangeRecall(s^*)$ on $\CountCutoff(s^*)$ separately.
This greatly simplifies calibrating the model,
avoiding the need to grid-search over possible pairs of both parameters.
Given targeted correctness of $s^*$, we determine $\ScoreCutoff(s^*)$
corresponding to desired $\TestRecall(s^*)$ based on
\autoref{fig:baseline-score-cutoff} and independently $\CountCutoff(s^*)$
corresponding to desired $\ChangeRecall(s^*)$ based on
\autoref{fig:baseline-count-cutoff}.
While, in principle, either recall depends on the choice of \emph{both}
$\ScoreCutoff(s^*)$ and $\CountCutoff(s^*)$, we have found that in order to
achieve high $\TestRecall(s^*)$ while $\ScoreCutoff(s^*) = 0$ one would have to
set $\CountCutoff(s^*)$ to large value, which would cause all possibly affected
test targets to run on large portion of changes.
Likewise, in order to achieve high $\ChangeRecall(s^*)$ while $\CountCutoff(s^*) =
0$ one would have to set $\ScoreCutoff(s^*)$ such that tests with very low
probability of failing are exercised, which too increases the number of selected
test targets.

At Facebook, we calibrate the test selection model $s^*$ to guarantee
$\TestRecall(s^*) > 0.95$ and $\ChangeRecall(s^*) > 0.999$ at land-time stage of
testing, as marked on \autoref{fig:baseline-score-cutoff} and
\autoref{fig:baseline-count-cutoff}.
As a consequence, we fail to report only $< 5\%$ of individual test failures and $<
0.1\%$ of faulty changes.
Based on our experience operating such test selection strategy in production for
several months, the described correctness guarantees are sufficient.
Note that any faulty change that makes it into the master branch will be
detected in the stabilization stage.
Besides, significantly more faults are detected in stabilization stage due to
reasons discussed in \autoref{sec:land-time}, than due to the test selection
missing failing tests.

As seen in \autoref{fig:baseline-score-cutoff}, had the model only been
selecting $\LikelyFailing(s^*, d)$ test targets for each change $d$,
we would observe $\SelectionRate(s^*) < 0.25$.
\autoref{fig:baseline-count-cutoff} shows that had the model only been selecting
$\HighlyRanked(s^*, d)$ test targets for each change $d$,
it would choose no more than $\CountCutoff(s^*) = 230$ targets per change.
Note that the average number of targets transitively depending on files modified
in a change $d$ is $|\DependentTests(d)| \gg 1000$,
that is multiple times larger than $\CountCutoff(s^*)$.
Overall, combining both approaches, that is selecting $\SelectedTests(s^*,
d) = \LikelyFailing(s^*, d) \cup \HighlyRanked(s^*, d)$ for each change $d$,
yields a model that achieves $\TestRecall(s^*) > 0.95$,
$\ChangeRecall(s^*) > 0.999$ and $\SelectionRate(s^*) < 0.33$.

The impact of the described test selection on the scalability of the continuous
integration system can be best measured by the fact that deploying it has
reduced total number of test executions by a factor of three and total
infrastructure cost of testing code changes, measured in number of machines,
by a factor of two, relative to test selection based on build dependencies.

\subsection{Feature Selection}

The model with all the features defined in \autoref{sec:feature-engineering}
will not necessarily perform better than a model with a subset of them and we
therefore need to apply feature selection.
In order to evaluate feature importance we used a wrapper method
\cite{Guyon2003}:
for every feature above we evaluated the model on a full feature set and a full
set without the evaluated feature.
To measure the impact of a feature on a model we use ratios of
$\SelectionRate(s^*)$ given $\TestRecall(s^*) = 0.9$ for the classification
metric and $\CountCutoff(s^*)$ ratios for the ranking metric.
\autoref{tab:feature-performance} summarizes performance improvements and
regressions for the features defined in \autoref{sec:feature-engineering},
the higher value is associated with better performance.
Values below 1 indicate a regression that was introduced by the feature.

\begin{table}[ht]
\renewcommand{\arraystretch}{1.25}
\caption{Relative feature performance}
\label{tab:feature-performance}
\centering
\begin{tabular}{c|c|c}
\hline
\bfseries Feature & \bfseries Classification & \bfseries Ranking
\\\hline\hline
File extensions & 1.04 & 1.62
\\\hline
Change history for files & 1.03 & 1.59
\\\hline
File cardinality & 0.95 & 0.98
\\\hline
Target cardinality & 1.1 & 0.2
\\\hline
Historical failure rates & 1.37 & 1.62
\\\hline
Project name & 1.15 & 0.97
\\\hline
Number of tests & 1.07 & 2.89
\\\hline
Minimal distance & 1.23 & 0.96
\\\hline
Common tokens & 0.33 & 0.68
\\\hline
Distinct authors & 0.3 & 0.72
\\\hline
\end{tabular}
\end{table}

The best performing model uses \emph{file extensions},
\emph{change history}, \emph{failure rates},
\emph{project name}, \emph{number of tests} and \emph{minimal distance}.
The remaining features introduce regressions and we excluded them from the
models used in \autoref{sec:model-deployment}.
Despite the regression on the ranking metric,
we include the \emph{project name} feature.
The feature improves the classification metric,
which dominates selection rate of the strategy.

\subsection{Impact of Test Flakiness}

\begin{figure}[ht]
\includegraphics[clip,trim=0 0.25cm 0 1cm,width=\linewidth]{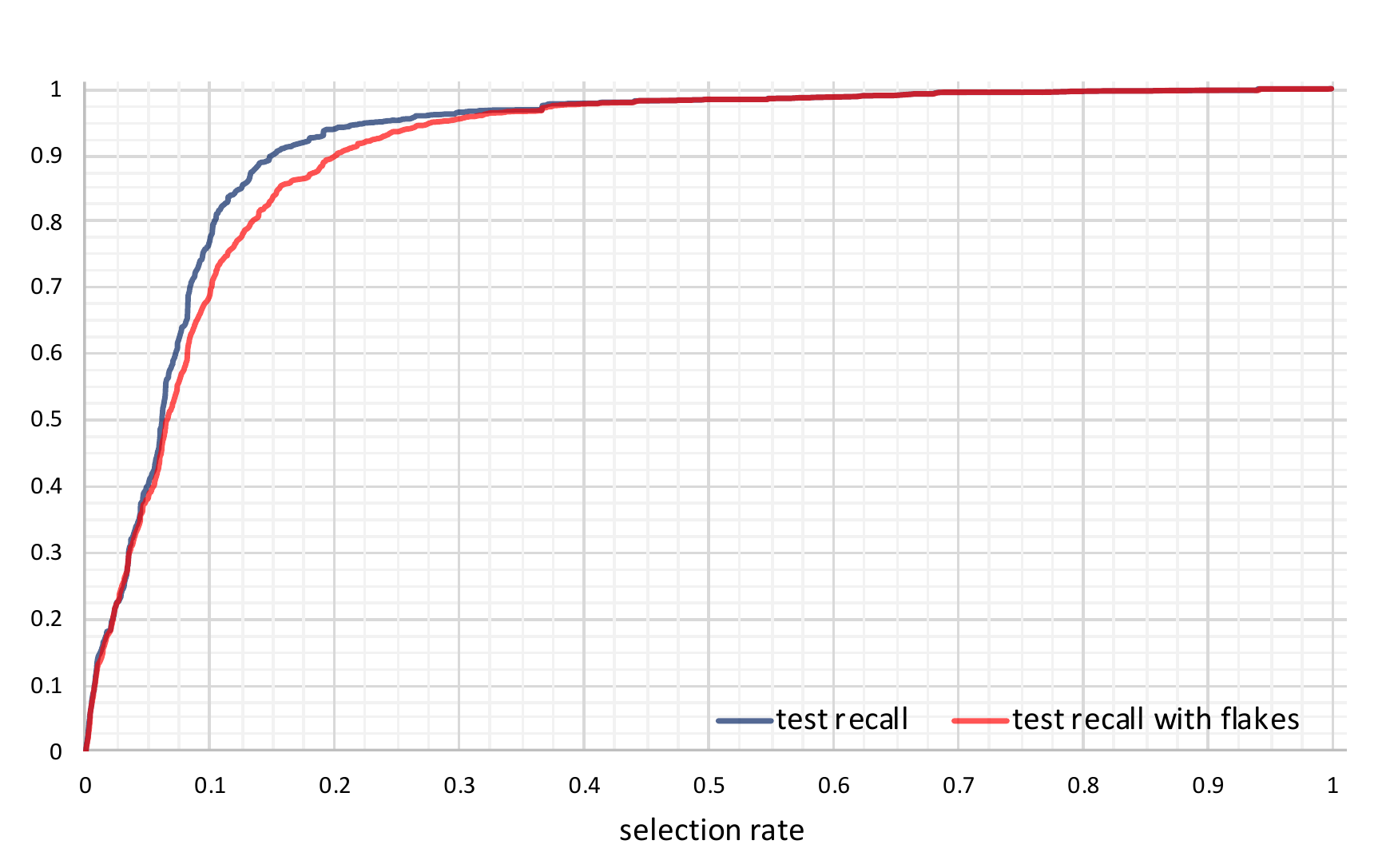}
\caption{
  $\TestRecall(s_A)$, $\TestRecallWithFlakes(s_A)$ as functions of $\SelectionRate(s_A)$
  for $\CountCutoff(s_A) = 0$ in Experiment A.
  The model is better at ``catching'' failed tests than those that would flake,
  as indicated by the fact $\TestRecall(s_A) \ge \TestRecallWithFlakes(s_A)$.
  \label{fig:flakiness-train-on-real}
}
\vspace{\baselineskip}
\includegraphics[clip,trim=0 0.25cm 0 1cm,width=\linewidth]{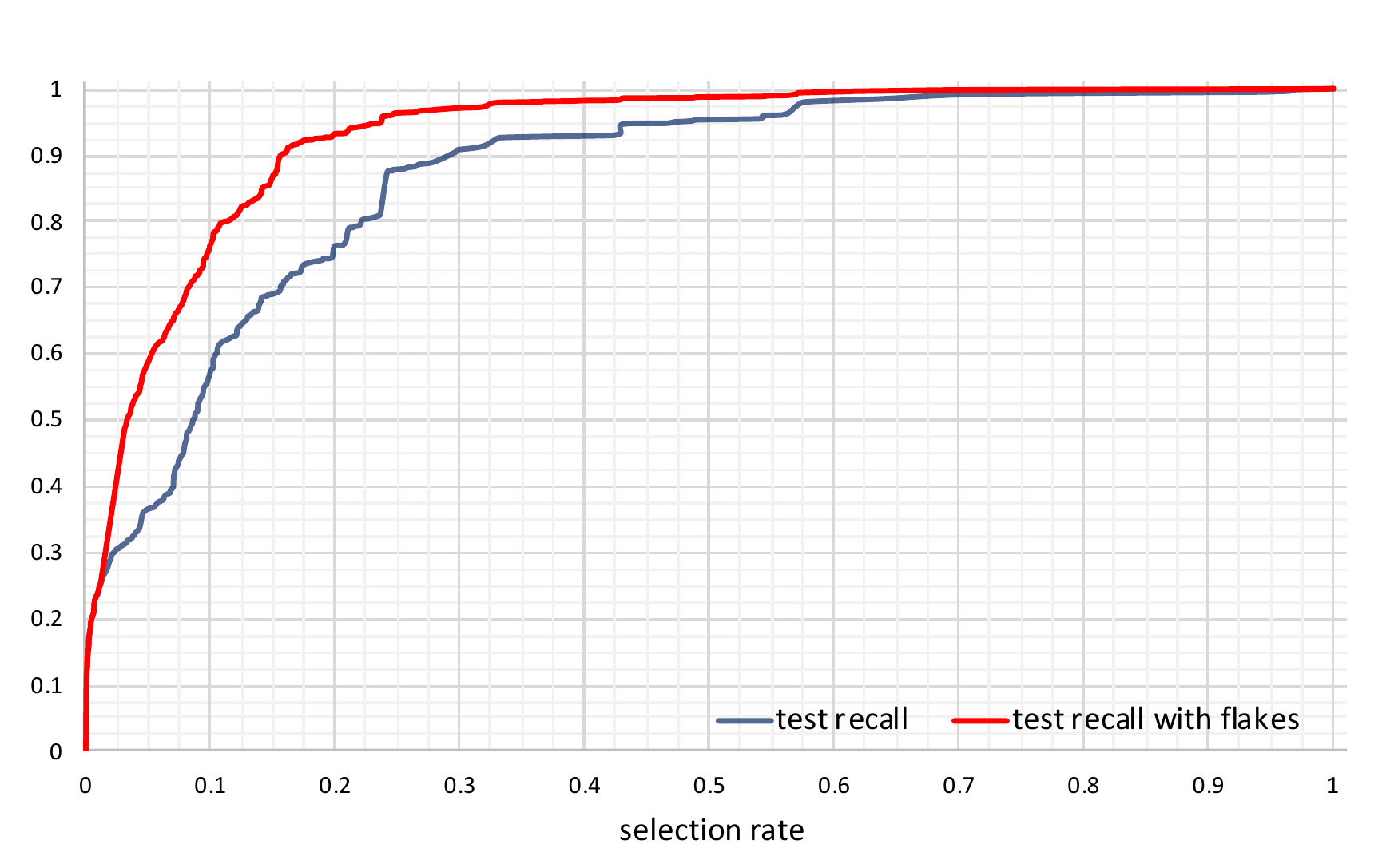}
\caption{
  $\TestRecall(s_B)$, $\TestRecallWithFlakes(s_B)$ as functions of $\SelectionRate(s_B)$
  for $\CountCutoff(s_B) = 0$ in Experiment B.
  The model trained and calibrated on non-de-flaked data using $\TestRecallWithFlakes(s_A)$
  would achieve worse-than-expected $\TestRecall(s_A)$,
  thus failing to report larger-than-expected number of test failures,
  as indicated by the fact $\TestRecall(s_B) < \TestRecallWithFlakes(s_B)$.
  \label{fig:flakiness-train-on-all}
}
\end{figure}

We have shown in \autoref{sec:test-flakiness} a theoretical argument that an
ability to identify failures unrelated to a code change is important when
evaluating empirical performance of a test selection strategy.
An interesting question is whether flakiness impacts learned test selection
models \emph{in practice}.
To answer it, we have conducted additional experiments that measure performance
of a model trained on data that did not undergo the described de-flaking
procedure.

In order to quantify to what extent a test selection strategy captures failures
unrelated to the underlying change, we define a measure similar to
$\TestRecall$.
\begin{definition}[Test recall with flakes]

  Let $s$ be a test selection strategy and $D$ a set of code changes,
  such that for $F_d = \FailedTests(d) \cup \FlakedTests(d)$,
  $\exists_{d \in D} F_d \neq \emptyset$.
  \begin{equation*}
    \TestRecallWithFlakes(s, D) = \frac{
      \sum_{d \in D} {|\SelectedTests(s, d) \cap F_d|}
    }{
      \sum_{d \in D} {|F_d|}
    }
  \end{equation*}

\end{definition}

Intuitively, the new metric equals empirical probability of a particular test
selection strategy ``catching'' a failed test outcome irrespective of whether it
was related to the change or not.
In the following experiments, we compare values of $\TestRecall(s)$ and
$\TestRecallWithFlakes(s)$ to determine whether the model is learning to
identify tests that would detect a fault or those that would flake.

In Experiment A, we have trained a test selection model $s_A$ as described in
\autoref{sec:test-selection-model}.
Using the evaluation dataset, we have plotted $\TestRecall(s_A)$,
$\TestRecallWithFlakes(s_A)$ as functions of $\SelectionRate(s_A)$ for
$\CountCutoff(s_A) = 0$ as seen in \autoref{fig:flakiness-train-on-real}.

In Experiment B, we have trained the test selection model $s_B$ as described in
\autoref{sec:test-selection-model}, with one modification.
This time, when training the binary classifier described in
\autoref{sec:model-architecture}, we considered examples $(d,
t)$ for $t \in \DependentTests(d)$ and $d \in D$ as positives if and only if $t
\in \FailedTests(d) \cup \FlakedTests(d)$.
This is equivalent to training the model on test outcomes recorded by
hypothetical learning test runs that did not perform aggressive retries
described in \autoref{sec:test-flakiness}.
Using the evaluation dataset, we have plotted $\TestRecall(s_B)$,
$\TestRecallWithFlakes(s_B)$ as functions of $\SelectionRate(s_B)$ for
$\CountCutoff(s_B) = 0$ as seen in \autoref{fig:flakiness-train-on-all}.

A number of observations based on the presented results of the experiments lead
to interesting conclusions.
\begin{enumerate}

  \item We have $\TestRecall(s_B) < \TestRecallWithFlakes(s_B)$ for all
  $\SelectionRate(s_B) \in [0, 1]$.
  Had we not performed the de-flaking procedure described in
  \autoref{sec:test-flakiness}, we would train and evaluate test selection model
  on a dataset that conflates failed and flaked tests.
  As a result, we would perceive the model to capture
  $\TestRecallWithFlakes(s_B)$ fraction of failures at chosen selection rate.
  In reality the model would capture only $\TestRecall(s_B)$ of failures at this
  selection rate.
  Note that at $\SelectionRate(s_B) = 0.15$ we have $\TestRecallWithFlakes(s_B)
  \approx 0.9$ but $\TestRecall(s_B) \approx 0.7$.
  Had we deployed such a model in production, we would fail to report
  \emph{three times as many test failures} as expected from the evaluation.

  \item For all $\SelectionRate(s_A) = \SelectionRate(s_B) \in [0, 1]$ we have
  $\TestRecall(s_A) > \TestRecall(s_B)$.
  This confirms that training the model on data that did not go through
  de-flaking procedure yields a model with strictly worse performance than had
  training data been de-flaked.

  \item $\TestRecall(s_A) \ge \TestRecallWithFlakes(s_A)$ for all choices of
  $\SelectionRate(s_A) \in [0, 1]$.
  This verifies that the model trained on de-flaked data is not worse at
  ``catching'' failed tests than those that flaked, a desired behavior.

\end{enumerate}

We can therefore conclude that it is important to reduce the impact of flakiness
on data used \emph{both for training and evaluation},
to prevent the model from learning to capture mostly flaky failures as well as
to be able to accurately measure its performance.

\section{Related Work} \label{sec:related-work}

A number of test selection techniques based on \emph{static analysis} of source
code at varying granularity have been proposed to date.
Ryder and Tip \cite{Ryder2001} present a test selection strategy based on
method-level analysis of call graphs.
Legunsen et al.
\cite{Legunsen2016, Legunsen2017} conducted an extensive study of static
techniques, noting that ones based on analyzing class-level test dependencies
match performance of state-of-art dynamic methods,
such as Ekstazi \cite{Gligoric2015}.
Zhang \cite{Zhang2018} described a test selection strategy that combines method-
and file-level analysis of test dependency and change information.
The mentioned techniques are not easily extensible to multilingual code bases,
where control flow of a program can cross language boundaries.
Also, analyzing code dependencies at fine granularity poses scaling challenges
in multi-million line code bases.

On the other hand, we are aware of multiple test selection strategies based on
\emph{dynamic analysis}.
Rothermel et al.
\cite{Rothermel1997} first describe a dynamic test selection technique operating
at a granularity of basic blocks in control flow graph.
Gligoric et al.
\cite{Gligoric2015} proposed a method operating at the granularity of files.
Celik et al.
\cite{Celik2017} present a technique capable of tracing test execution across
language boundaries.
The mentioned dynamic techniques require recording execution traces at
sufficiently fine granularity, which is not feasible at Facebook's scale.

Memon et al.
\cite{Memon2017} describe a technique most closely related to our work,
coming from similar industrial context.
The technique has been applied in large, monolithic code base and combines
static analysis of build metadata with the empirical observation that changed
units of code and failing tests have small distance in build dependency graph.
In our predictive test selection strategy, we use the distance as one of the
features and find that, although important, it is not sufficient to conduct
accurate test selection on its own.

\section{Conclusions and Future Directions}
\label{sec:conclusions-and-future-directions}

Delivering test signal to engineers early in the development workflow is crucial
to developer productivity.
Continuous integration systems must, however,
balance the quality of signal with its latency and cost,
which can be achieved through change-based test selection.
Designing and implementing scalable test selection strategy is a non-trivial
problem, especially in large monolithic code bases.
We have demonstrated that such a strategy can in fact be \emph{learned}
automatically from a sufficiently large dataset containing outcomes of tests
exercised on various code changes.
By combining many sources of information, each being a weak indicator of whether
a particular test needs to be run on a specific version of code on its own,
we were able to construct a test selection strategy delivering good and
predictable performance.
Applying machine learning techniques allowed us to maintain the strategy with
little to no manual tuning that is typically necessary with various heuristics.
We have also shown that non-determinism of real-world tests does not preclude
applicability of predictive test selection.

Many important sources of information, such as history of code changes,
could be incorporated into the test selection model in the form of additional
features.
We have shown that the model effectively combines multiple seemingly unrelated
pieces of information in order to assess the probability of a particular test
failing.
We believe that existing change-impact analysis techniques can yield many more
powerfull features we did not have a chance to explore in this work.

We are also interested in experimenting with more sophisticated machine learning
algorithms and model architectures.
The presented approach considers each test potentially impacted by a change
separately, thus it cannot capture the fact that some subsets of tests may have
overlapping coverage and thus correlated results.
We believe it is possible to make the predictive test selection strategy
understand correlations between tests outcomes and avoid selecting multiple sets
that are likely going to provide redundant signal \cite{Najafi2019}.

\section*{Acknowledgments}
\label{sec:acknowledgments}

We would like to thank the following engineers and acknowledge their
contributions to the project: Billy Cromb, Jakub Grzmiel, Glenn Hope, Hamed
Neshat, Andrew Pierce, Yuguang Tong, Eric Williamson, and Austin Young.

\vspace{2\baselineskip}
\bibliographystyle{IEEEtran}
\bibliography{IEEEabrv,main}

\end{document}